# Visibility study of graphene multilayer structures

Guoquan Teo and Haomin Wang

*Department of Electrical and Computer Engineering, National University of Singapore, 4 Engineering Drive 3, Singapore 117576 and Data Storage Institute, 5 Engineering Drive 1, Singapore 117608*

Yihong Wu [*]

*Department of Electrical and Computer Engineering, National University of Singapore, 4 Engineering Drive 3, Singapore 117576*

Zaibing Guo and Jun Zhang

*Data Storage Institute, 5 Engineering Drive 1, Singapore 117608*

Zhenhua Ni and Zexiang Shen

*Division of Physics and Applied Physics, School of Physical and Mathematical Sciences, Nanyang Technological University, 1 Nanyang Walk, Block 5, Level 3, Singapore 637616*

---

[*] Author to whom correspondence should be addressed; electronic mail: elewuyh@nus.edu.sg

**ABSTRACT:**

The visibility of graphene sheets on different types of substrates has been investigated both theoretically and experimentally. Although single layer graphene is observable on various types of dielectrics under an optical microscope, it is invisible when it is placed directly on most of the semiconductor and metallic substrates. We show that coating of a resist layer with optimum thickness is an effective way to enhance the contrast of graphene on various types of substrates and makes single layer graphene visible on most semiconductor and metallic substrates. Experiments have been performed to verify the results on quartz and NiFe-coated Si substrates. The results obtained will be useful for fabricating graphene-based devices on various types of substrates for electronics, spintronics and optoelectronics applications.

## I. INTRODUCTION

The recent success in extracting a single layer of carbon from highly ordered pyrolytic graphite (HOPG) using a technique called micromechanical cleavage has stimulated a large interest in both the fundamental properties and potential applications of graphene.[1,2] Although graphene can also be formed epitaxially on SiC[3] through thermal annealing, currently mechanical exfoliation of graphite is still the most efficient way to obtain high-quality graphene samples. As it is reported in literature[4], mechanical exfoliation of graphite usually results in single layer graphene (SLG), bilayer graphene (BLG) and few layer graphene (FLG) sheets. Therefore, the first step to fabrication of graphene-based devices is to identify graphene flakes with different number of layers and determine their relative positions on the wafer with respect to the pre-formed alignment marks. In order to make this first step happen one must find a way to make the graphene visible under an optical microscope. Although graphene sheets can be seen by atomic force, scanning electron and optical microscopes, the first two techniques cause either damage or contamination to the graphene sheets.[1,2] Therefore, currently only optical microscopy allows a non-destructive recognition of graphene sheets. The advantage of optical microscopy is that, in addition to optical contrast, Raman spectra are also helpful in determining the thickness of graphene sheets So far, several papers have been published on visibility of graphene on $SiO_2$ coated Si substrates with or without a resist.[5-11] From application point of view, however, it is also of interest to know if graphene can be "seen" on other types of substrates or layers, in particular, those that are widely used for existing or will be potentially used for future electronic, optoelectronic and spintronic devices. Based on this background, in this paper we report on the visibility study of SLG, BLG and FLG on a wide range of metal, semiconductor and oxide layers or substrates. We show by simulation that the contrast of graphene can be enhanced substantially with the addition of a resist layer with optimum thickness on almost all types of substrates. This will make SLG visible on most semiconductor and metallic substrates which is otherwise unobservable under an optical microscope. Experiments have been performed to verify the results on quartz and NiFe-coated Si substrates.

## II. THEORETICAL MODEL

The visibility of graphene on different types of substrates with or without under/capping layer originates from both the relative phase shift and amplitude modification induced by the graphene layer. Although graphene is only one atomic layer thick, its refractive index was found to be very close to that of graphite. Therefore, the reflection spectrum of a multiple layer structure containing graphene can be readily obtained using the 2×2 matrix method.[12] Fig. 1 shows a typical multilayer structure consisting of air, N layers of homogeneous media and supporting substrate. The thickness and refractive index of each layer are $d_0$, $d_i$ (i = 1 to N), $d_S$ and $n_0$, $n_i$ (i = 1 to N), $n_S$, respectively. Here, the reflective indices are in general complex numbers. We also assume that $d_0$ and $d_s$ approaches infinity. Assuming that electromagnetic wave travels in zx plane and the media are homogeneous in z-direction, the electrical field that satisfies the Maxwell equation can be written as $E = [A(x) + B(x)]e^{i(\omega t - \beta z)}$, where β is the z component of the wave vector, ω is the angular frequency, t is time, and A(x) and B(x) are amplitude of the right-travelling and left-travelling waves, respectively. The electromagnetic wave can be either a p-wave or an s-wave. The amplitude of the electrical field inside the air and those after passing through the Nth layer and substrate interface are related by the following equation:

$$\begin{pmatrix} A_0 \\ B_0 \end{pmatrix} = D_0^{-1} \left[ \prod_{m=1}^{N} D_m P_m D_m^{-1} \right] D_S \begin{pmatrix} A_S \\ B_S \end{pmatrix},$$

where

$$D_m = \begin{pmatrix} 1 & 1 \\ n_m \cos\phi_m & -n_m \cos\phi_m \end{pmatrix} \quad \text{for s wave}$$

$$D_m = \begin{pmatrix} \cos\phi_m & \cos\phi_m \\ n_m & -n_m \end{pmatrix} \quad \text{for p wave}$$

$$P_m = \begin{pmatrix} e^{i\omega_m} & 0 \\ 0 & e^{-i\omega_m} \end{pmatrix}$$

$$\omega_m = \frac{2\pi n_m d_m}{\lambda}$$

Here, $n_m$ is the refractive index of the *m*th layer, $d_m$ is the thickness of *m*th layer, $\lambda$ is wavelength, and $\phi_k$ is angle of incidence. If we let

$$D_0^{-1} \left[ \prod_{m=1}^{N} D_m P_m D_m^{-1} \right] D_S = \begin{pmatrix} T_{11} & T_{12} \\ T_{21} & T_{22} \end{pmatrix},$$

then the reflectance is given by

$$R = \left| \frac{T_{21}}{T_{11}} \right|^2 .$$

For unpolarized light, one can take an average of the contributions from both the *s* wave and *p* wave. The contrast C induced by the graphene layer is then given by

$$C = \frac{R(\text{no\_graphene}) - R(\text{with\_graphene})}{R(\text{no\_graphene})} .$$

The simulation has been carried out using Matlab. The incident wave is assumed to be perpendicular to the plane of the multiple layers; hence the angle of incident $\phi_k$ was set at zero degree. The complex refractive indices of all materials used in the paper are adopted from literature[13] except ZnO. Single layer graphene is assumed to have a thickness 0.34 nm, and multilayer graphene which consists of n monolayers is assumed to have a thickness of n×0.34 nm. The refractive index of graphene is assumed to be the same as that of bulk graphite and is independent of $\lambda$, i.e., $n_G(\lambda) = 2.6-1.3i$.

## III. RESULTS AND DISCUSSION

### A. Contrast spectra of graphene sheets on Au, Si and $Al_2O_3$

The simplest structure involving graphene is a layer of SLG or FLG on a flat substrate, as shown schematically in Fig. 2(a). As representative examples of, oxides, semiconductor and metal, we have calculated the contrast spectra of graphene sheets with different thicknesses on sapphire ($Al_2O_3$), silicon (Si) and gold (Au). The results are shown in Figs. 2(b), (c) and (d), respectively, for graphene sheets with number of layers ranging from 1 to 10. As can be seen from the figures, graphene on $Al_2O_3$ and Si exhibits a negative contrast while graphene on Au shows a positive one. In other words, graphene on $Al_2O_3$ and Si appears brighter than the substrate while graphene on Au is darker than the substrate. From Figs. 2(b), (c) and (d), it is observed that the visibility for single layer graphene on gold or silicon (contrast of 0.0270 for gold and -0.0066 for silicon) is not as good as that for single layer graphene on a sapphire substrate (contrast of -0.0664) and is practically difficult to be seen under an optical microscope. On the other hand, single layer graphene sheets on sapphire are not difficult to be seen because the contrast is close to 0.09, almost the same as that for SLG on $SiO_2$ (300 nm) / Si structure.[6,9] As both the absorption and phase modification contribute to the contrast formation mechanism, the contrast generally increases with the number of layers, as shown in Figs. 2(b)-(d).

### B. Contrast spectra of PMMA / graphene / substrate structures

As it has been reported in literature, the contrast for single layer graphene can be increased substantially by adding a layer either underneath or atop the graphene layer. The former also facilitate the fabrication of field-effect devices if the inserted layer can also function as a gate oxide like SiO$_2$ and HfO$_2$. On the other hand, the latter can be a resist layer, such as polymethylmethacrylate (PMMA), which is a commonly used resist for e-beam lithography. The schematic of such a structure is shown in Fig. 3. The advantage of using PMMA over-layer is twofold. One of the advantages is that the thickness of PMMA can be controlled precisely through optimizing the resist solution and its coating process. In addition to adjustable thicknesses, the resist can also be used to pattern the graphene for device fabrication.

Figs. 4(a) to 4(f) show the simulated contrast of graphene as a function of wavelength for PMMA thicknesses ranging from 0 to 400 nm on different types of oxide and nitride substrates: (a) SiO$_2$, (b) Si$_3$N$_4$, (c) HfO$_2$, (d) Al$_2$O$_3$, (e) MgO and (f) TiO$_2$. Although the contrast is not high when graphene along is placed on these substrates, the contrast is significantly enhanced with the addition of a PMMA overlayer, especially for SiO$_2$, Al$_2$O$_3$ and MgO substrates over a wide range of the spectrum. In the case of Si$_3$N$_4$, HfO$_2$ and TiO$_2$ substrates, viewing of graphene under the microscope would probably require the use of a narrow band pass filter as contrast of graphene is greatly enhanced only over a narrow band of the spectrum. It is also noted that, in most cases, positive contrasts are obtained, in which graphene is darker than the background and also more importantly, the contrasts are comparable, if not better than the current graphene-SiO$_2$-Si structure for identifying graphene.

Figs. 5(a) to 5(f) show the simulated contrast of graphene as a function of wavelength for PMMA thicknesses ranging from 0 to 400 nm on different types of semiconductor substrates: (a) Si, (b) Ge, (c) GaAs, (d) GaN, (e) ZnO and (f) ZnSe substrate. Although improvements have been obtained in all cases, only the contrast on GaN, ZnO and ZnSe has been increased to a level that is comparable to that of graphene-SiO$_2$-Si structure. As it is in the cases of Si$_3$N$_4$, HfO$_2$ and TiO$_2$ substrates, the contrast enhancement for GaN and ZnO substrates is obtained only in a narrow band of the spectrum; therefore, a narrow band pass filter is probably needed for viewing graphene under the microscope. Zinc oxide is the only semiconductor among the six that shows a positive contrast, and the rest has a negative contrast. In the case of Si, Ge and GaAs, it would probably still be possible to "see" graphene even though the contrast values are at the lower limit of the observable range of naked eyes.

Figs. 6(a) to 6(f) show the simulated contrast of graphene as a function of wavelength for PMMA thicknesses ranging from 0 to 400 nm on different types of metallic substrates: (a) Co, (b) Ni, (c) Fe, (d) NiFe, (e) Au and (f) Cu. It is worth noting that the contrasts for the metals have more or less doubled. Although the enhancement is significant, the absolute value of contrast is still rather low, which makes the identification of graphene a difficult task on metal substrates. Gold and copper are the two among the six metals having a higher contrast that could be observed easily under the microscope with the naked eyes.

In Table I, we summarize the maximum contrasts obtained on different types of substrates with or without a resist layer of optimum thickness and the corresponding enhancement ratio. Based on our experimental experience, the lower contrast limit to see graphene by naked eyes is about 0.02. Therefore, it is possible to see graphene under an optical microscope on most of the substrates after the coating of a PMMA layer with optimum thickness. As shown in Figs. 4 - 6, the optimum thickness for PMMA ranges from 50 to 100 nm in the visible wavelength range, the approach is technically feasible.

Fig. 7(a) to 7(d) shows some of the simulation results for thicker graphene layers (N = 1 to 4) on different types of substrates, with the thickness of PMMA optimized for maximum contrast: (a) silicon dioxide substrate with 275 nm of PMMA, (b) silicon substrate with 103 nm of PMMA, (c) gallium arsenide substrate with 308 nm of PMMA and (d) cobalt substrate with 310 nm of PMMA. It can be seen clearly that as the number of layers increases, the contrast increases linearly as well. Hence, few layer graphene sheets can be easily identified on various types of substrates and distinguished from single layer ones.

### C. Experimental verification

Experiments were performed to verify the simulation results on two representative substrates: quartz and NiFe (60 nm) coated Si wafer. The graphite flakes obtained by mechanical exfoliation were transferred to the substrates which include SLG, BLG and FLG sheets. Subsequently, the contrast and Raman spectra were measured to identify SLG and BLG sheets. After that, a thin layer (~ 380 nm for quartz and ~ 325 nm for NiFe ) of PMMA was coated on the substrates with graphene sheets, followed by the second round of optical measurements. Fig. 8 (a) compares the measured contrast spectra for a SLG and BLG with the simulation results on quartz substrate. The measured contrast is about half that of the simulated value in the entire visible wavelength range, which might be caused by the influence of reflection from the back surface. As shown in Fig. 8(b), the coating of a 380 nm PMMA enhances slightly the contrast in the wavelength range of 466 nm – 616 nm, but changes the contrast from negative to positive in the range of 550 nm – 750 nm. The experimental data are in good agreement with the simulation results. In the case of NiFe, which is the most commonly used material for spintronics, we could not obtain any contrast when it was not coated with PMMA. However, as shown in Fig. 8(c), a reasonably high contrast is obtained for samples coated with a PMMA layer near the blue wavelength region for both SLG and BLG. Again, a good agreement is obtained between simulated and measured data. SLG became visible under optical microscope, though it is not as clear as when it is on the quartz substrate. Therefore, PMMA coating is a technically feasible way to enhance graphene contrast on different types of substrates.

## IV. SUMMARY

In summary, we have studied theoretically the visibility of graphene on different types of substrates with or without a resist layer. These materials include oxides, nitride, semiconductors and metals. It was found that SLG can be directly observed on oxide and nitride while it is invisible on most semiconductors and metals by optical microscope. The coating of a PMMA layer with optimum thickness is an effective way to enhance the contrast of SLG on all types of substrates investigated and it also makes SLG visible on most semiconductors and metals. Experiments were performed to verify the results on quartz and NiFe/Si substrates. The results will be useful for fabricating graphene-based electronic, spintronic and optoelectronic devices on various types of substrates.

**ACKNOWLEDGEMENT**

The work at the National University of Singapore was supported by A*-STAR under Grant No. R-398-000-020-305.

**Figure Captions**

FIG. 1. A multilayer model used in the transfer matrix simulation.

FIG. 2. (a) Schematic of light reflection from a graphene sheet on a substrate; (b)-(d) Optical contrast spectra of graphene sheets with thickness ranging from 1 to 10 layers on: (b) sapphire substrate, (c) Si substrate and (d) Au substrate.

FIG. 3. Schematic of light reflection from a graphene sheet on a substrate coated with a PMMA overlayer.

FIG. 4. Contrast of graphene as a function of wavelength from 400 nm to 750 nm and PMMA thickness from 0 to 400 nm on different types of substrates or layers: (a) $SiO_2$, (b) $Si_3N_4$, (c) $HfO_2$, (d) sapphire, (e) MgO and (f) $TiO_2$.

FIG. 5. Contrast of graphene as a function of wavelength from 400 nm to 750 nm and PMMA thickness from 0 to 400 nm on different types of semiconductor substrates: (a) Si, (b) Ge, (c) GaAs, (d) GaN, (e) ZnO, (f) ZnSe.
.
FIG. 6. Contrast of graphene as a function of wavelength from 400 nm to 750 nm and PMMA thickness from 0 to 400 nm on different types of metallic substrates or thin films: (a) Co, (b) Ni, (c) Fe, (d) NiFe, (e) Au, (f) Cu.

FIG. 7. Contrast of graphene as a function of wavelength on (a) silicon dioxide substrate with 275 nm of PMMA, (b) Si substrate with 103 nm of PMMA, (c) GaAs substrate with 308 nm of

PMMA, and (d) Co with 310 nm of PMMA (d). The arrows indicate that the number of graphene layers increases from 1 to 4.

FIG. 8. Experimental and simulation results of contrast spectra of SLG and BLG for (a) graphene/quartz, (b) PMMA (380 nm)/graphene/quartz, and (c) PMMA (325 nm)/graphene / NiFe (60 nm) /Si.

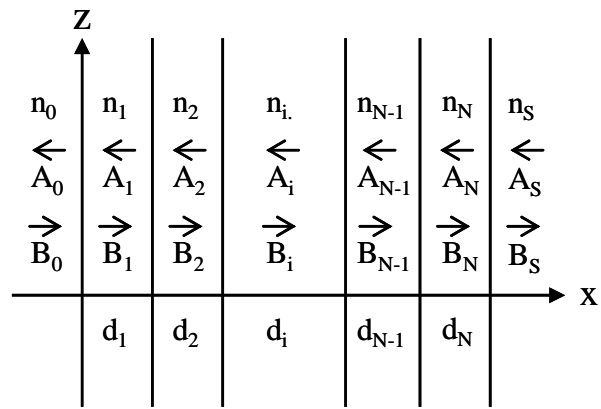

FIG. 1

GQ Teo et al.

J. Appl. Phys.

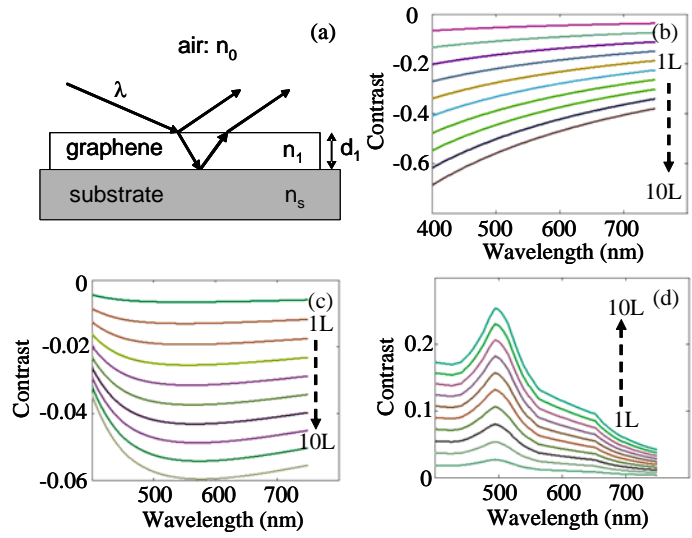

FIG. 2

GQ Teo et al.

J. Appl. Phys.

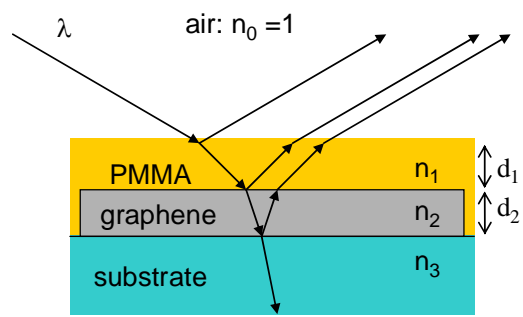

FIG. 3

GQ Teo et al.

J. Appl. Phys.

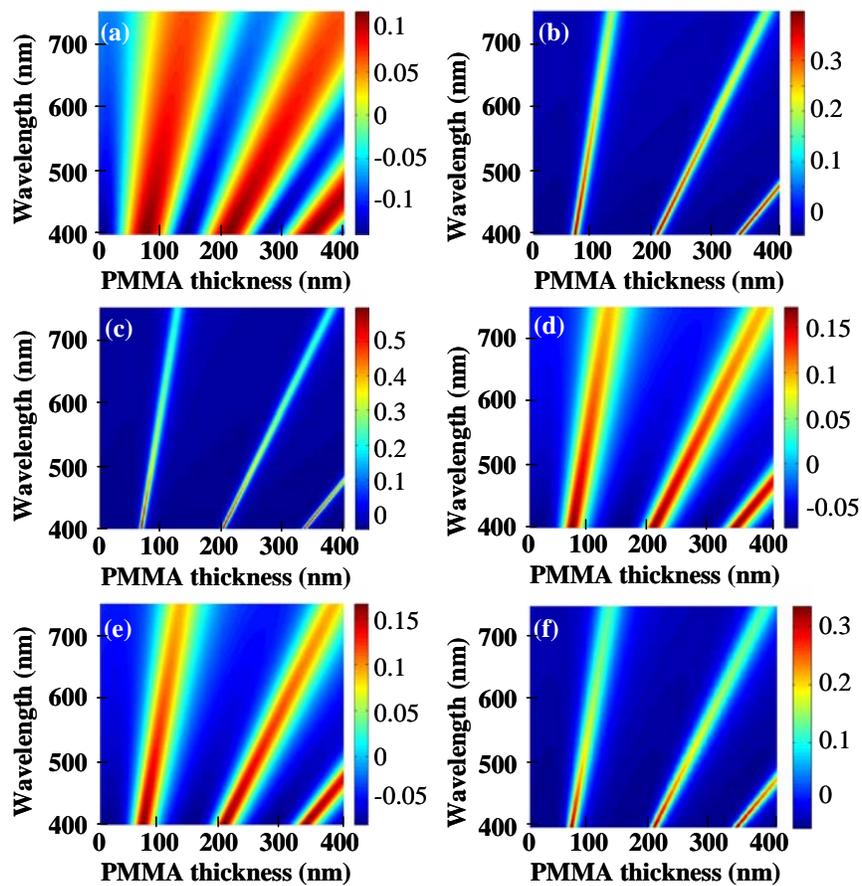

FIG. 4

GQ Teo et al.

J. Appl. Phys.

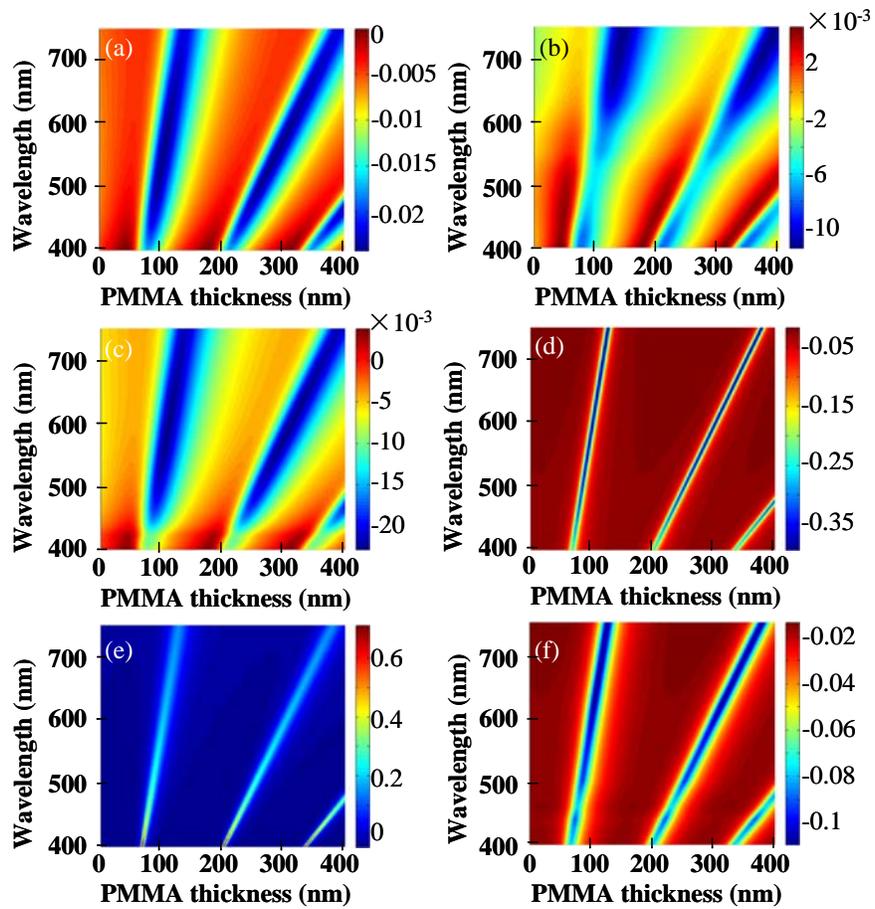

FIG. 5

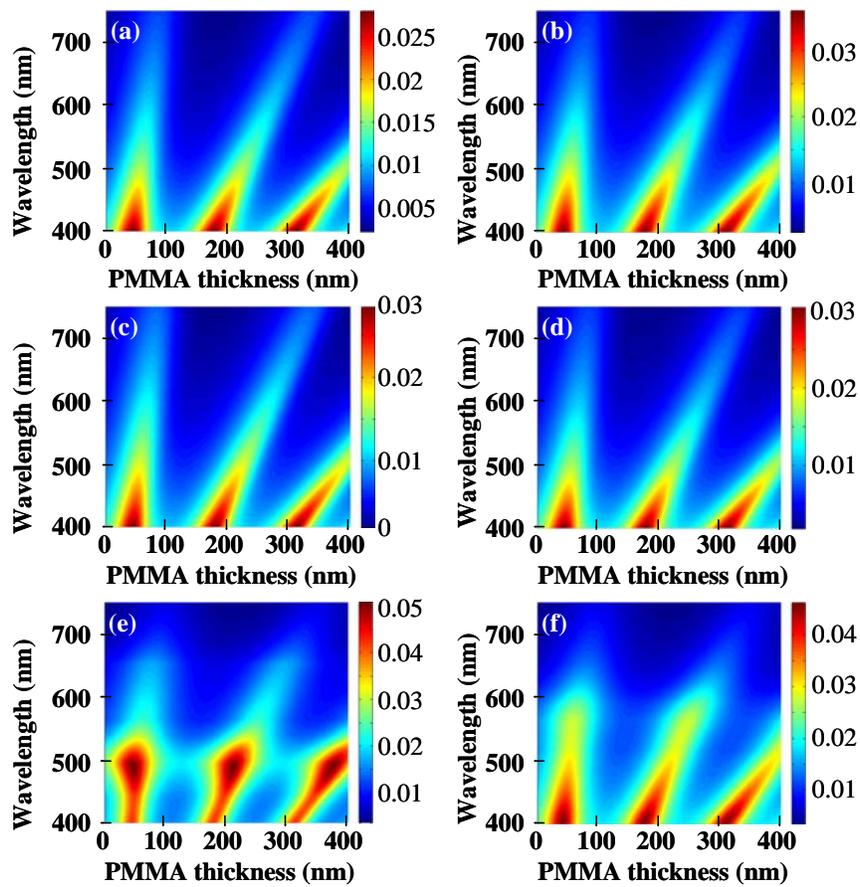

FIG. 6

GQ Teo et al.

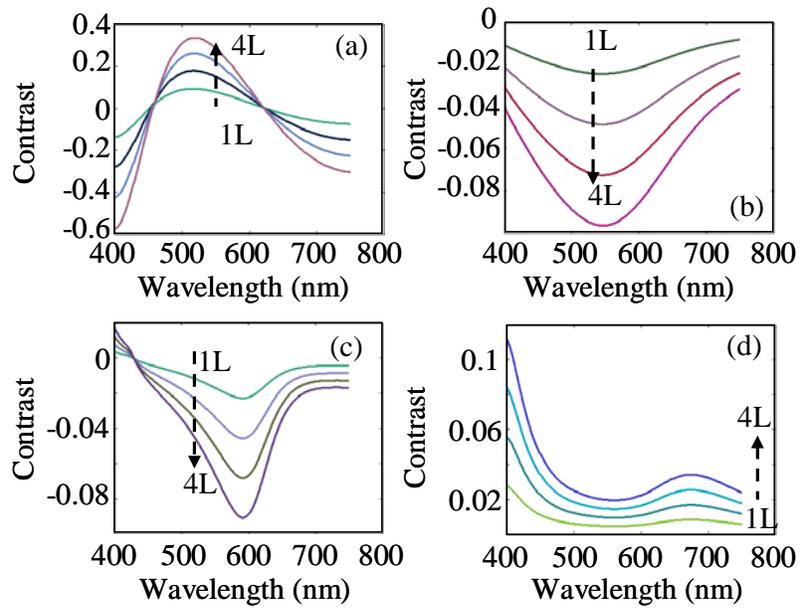

FIG. 7

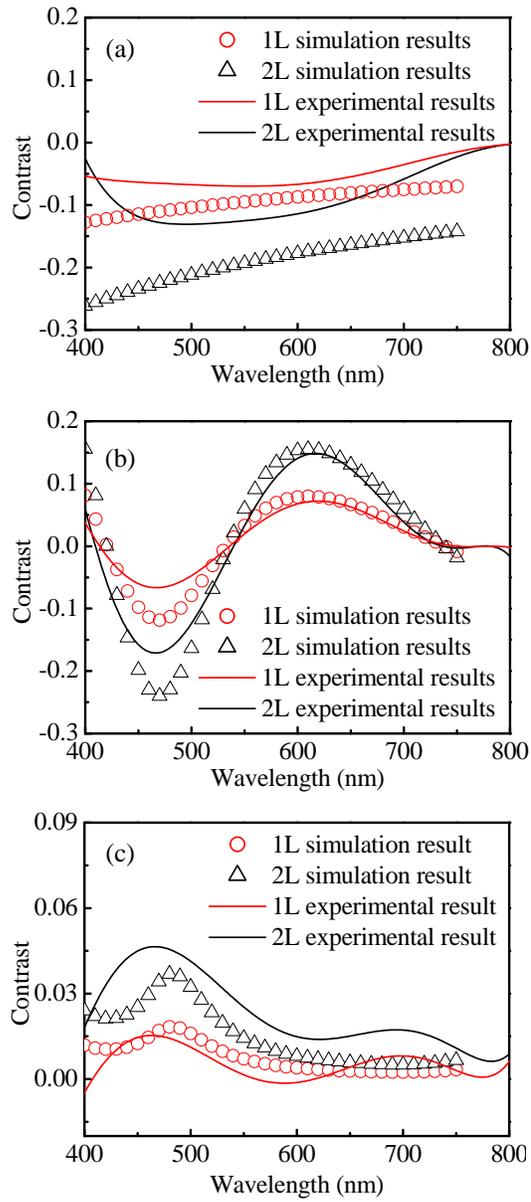

FIG. 8

GQ Teo et al.

J. Appl. Phys.

Table I. Summary of maximum contrasts obtained on different types of substrates with or without a resist layer of optimum thickness and the corresponding enhancement ratio.

| Type | Substrate | Max Contrast (w/o PMMA) | Max Contrast (w PMMA) | Thickness of PMMA (nm) | Improvement |
|---|---|---|---|---|---|
| Insulator | $SiO_2$ | -0.1277 | -0.1373 | 275 | 7.52% |
| | $Si_3N_4$ | -0.0438 | 0.3984 | 67 | 809.59% |
| | $HfO_2$ | -0.0403 | 0.5973 | 67 | 1382.13% |
| | $Al_2O_3$ | -0.0664 | 0.1722 | 69 | 159.34% |
| | MgO | -0.0692 | 0.1653 | 203 | 138.87% |
| | $TiO_2$ | -0.0456 | 0.3341 | 67 | 632.68% |
| TCO | ITO | -0.0457 | 0.3453 | 67 | 655.58% |
| Semiconductor | Si | -0.0066 | -0.0242 | 103 | 266.67% |
| | Ge | -0.0036 | -0.0140 | 398 | 288.89% |
| | GaAs | -0.0059 | -0.0229 | 308 | 288.14% |
| | GaN | -0.0266 | -0.3932 | 122 | 1378.20% |
| | ZnO | -0.0394 | 0.7101 | 67 | 1702.28% |
| | ZnSe | -0.0169 | -0.1096 | 313 | 548.52% |
| Metal | Co | 0.0134 | 0.0280 | 310 | 108.96% |
| | Ni | 0.0168 | 0.0359 | 310 | 113.69% |
| | Fe | 0.0130 | 0.0284 | 177 | 118.46% |
| | NiFe | 0.0147 | 0.0305 | 176 | 107.48% |
| | Au | 0.0270 | 0.0508 | 374 | 88.15% |
| | Cu | 0.0233 | 0.0462 | 40 | 98.28% |